\documentclass[prd,notitlepage,letterpaper,showpacs,preprintnumbers,amsmath,amssymb,nofootinbib,twocolumn, superscriptaddress, hyperref]{revtex4-1}
\usepackage{graphicx}
\usepackage{bm}
\usepackage[usenames,dvipsnames]{color}
\usepackage{hyperref}
\hypersetup{
    colorlinks=true,       
    linkcolor=NavyBlue,          
    citecolor=Magenta}   
\usepackage{soul,xcolor}

\newcommand\ee{\end{equation}}
\newcommand\be{\begin{equation}}
\newcommand\eea{\end{eqnarray}}
\newcommand\bea{\begin{eqnarray}}

\newcommand\ie{{\it i.e.}~}
\newcommand\eg{{\it e.g.}~}
\newcommand\del{\partial}
\newcommand\eq[1]{Eq.~(\ref{#1})}
\renewcommand\d[1]{\:\textrm{d}#1}

\renewcommand\({\left(}
\renewcommand\){\right)}
\renewcommand\[{\left[}
\renewcommand\]{\right]}

\begin{document}
\preprint{IFIC/15-10}

\setstcolor{red}

\title{Does Current Data Prefer a Non-minimally Coupled Inflaton?}
\author{Lotfi Boubekeur}
\affiliation{ Instituto de F\'isica Corpuscular (IFIC), CSIC-Universitat de Valencia,\\
Apartado de Correos 22085,  E-46071, Spain.}
\affiliation{ Laboratoire de Physique Math\'ematique et Subatomique (LPMS)\\
Universit\'e de Constantine I, Constantine 25000, Algeria.}

\author{Elena Giusarma}
\affiliation{Physics Department and INFN, Universit\`a di Roma ``La Sapienza'', Ple Aldo Moro 2, 00185, Rome, Italy}

\author{Olga Mena} 
\affiliation{
Instituto de F\'isica Corpuscular (IFIC), CSIC-Universitat de Valencia,\\ 
Apartado de Correos 22085,  E-46071, Spain.}

\author{H\'ector Ram\'irez}
\affiliation{ Instituto de F\'isica Corpuscular (IFIC), CSIC-Universitat de Valencia,\\
Apartado de Correos 22085,  E-46071, Spain.}
\begin{abstract}
{We examine the impact of a non-minimal coupling of the inflaton to the
Ricci scalar, $\frac12 \xi R\phi^2$, on the inflationary predictions. Such a non-minimal coupling is expected to be present in
the inflaton  Lagrangian on fairly general grounds. As a case study,
we focus on the  simplest inflationary model governed by the potential
$V\propto \phi^2$, using the latest combined 2015 analysis of {\it Planck} and BICEP2/\emph{Keck} Array. 
We find that the presence of a coupling $\xi$ is favoured at a
significance of $99\%$~CL, assuming that nature has chosen the potential
$V\propto \phi^2$ to generate the primordial perturbations and a number of e-foldings $N=60$. 
Within the context of the same scenario, we find that the value
of $\xi$ is different from zero at the $2\sigma$ level. When considering the cross-correlation polarization
spectra from BICEP2/\emph{Keck} Array and {\it Planck}, a value of
$r=0.038_{-0.030}^{+0.039}$ is predicted in this particular non-minimally coupled scenario. Future cosmological observations may therefore test these values of $r$ and verify or falsify the non-minimally coupled model explored here.
}

\end{abstract}

\maketitle

\twocolumngrid
\section{Introduction}

Inflation provides the most theoretically
attractive and observationally successful cosmological scenario able
to generate the initial conditions of our universe, while solving the
standard cosmological problems. Despite this remarkable success, the
inflationary paradigm is still lacking firm observational
confirmation. The picture that emerges from the latest data from {\it
  Planck}, including also the joint analysis of $B$-mode polarization
measurements from the BICEP2  collaboration \cite{Ade:2015tva,
  planckcosmo, planckng, planckinf}, is compatible with the
inflationary paradigm. According to these observations, structure
grows from Gaussian and adiabatic primordial perturbations.  From the
theoretical viewpoint, this picture is usually understood as the
dynamics of a single new scalar degree of freedom, {\it the inflaton},
minimally coupled to Einstein gravity.  However, the inflaton $\phi$
is expected to have  a non-minimal coupling to the  Ricci scalar
through the operator $\frac12\xi R \phi^2$, where $\xi$ is a
dimensionless coupling. Indeed, successful reheating requires that the
inflaton is coupled to the light degrees of freedom. Such couplings,
though weak, will induce a non-trivial running for $\xi$. Thus, even
starting from a vanishing value of $\xi$ (away from the conformal fixed point $\xi=-1/6$) at some energy scale, a non-trivial non-minimal coupling will be generated radiatively at some other scale (see \eg Ref.~\cite{Buchbinder:1992rb}). Therefore, it is important to study the impact of such a coupling on the inflationary predictions, especially in view of the latest {\it Planck}  2015 data.

Generically, for successful inflation, the inflaton should be very
weakly coupled\footnote{This requirement is also dictated by the
  non-detection of large primordial non-Gaussianities \cite{planckng}
  and the soft breaking of the shift symmetry $\phi\to\phi+c$, necessary to protect  the flatness of the potential.}. It follows that the magnitude of
$\xi$ is expected to be small.  Yet, even with such a suppressed
coupling, the  inflationary predictions are significantly
altered~\cite{Salopek:1988qh,Futamase:1987ua,Fakir:1990eg,Kaiser:1994vs,Komatsu:1999mt,Hertzberg:2010dc,Okada:2010jf,Linde:2011nh,Kaiser:2013sna,Chiba:2014sva,Pallis:2014cda}. For
instance, and as we will see, a small and positive $\xi$ can enlarge
considerably the space of phenomenologically acceptable scenarios (see also \cite{Tsujikawa:2013ila}). In
this paper, we will focus on the simplest inflationary
scenario with a potential $V\propto \phi^2$~\cite{Linde:1983gd}, and a
non-zero non-minimal coupling. According to the very recent {\it
  Planck} 2015 full mission results, the minimally-coupled version of
this scenario (\ie $\xi=0$) is ruled out at more than $99\%$
confidence level~\cite{planckcosmo, planckinf}, for $50$ e-folds of
inflation. Nevertheless, the $N=60$ case is only moderately disfavoured at
$95\%$~ CL. Thus, before discarding it definitely from the range of
theoretical possibilities, it is worthwhile to explore this scenario
in all generality (considering as well different posibilities for the
number of e-folds), given
that, as explained earlier, the presence of non-minimal couplings in
the inflaton Lagrangian is quite generic.

\section{Non-minimally coupled Inflaton} 

The dynamics of a non-minimally coupled scalar field $\phi$
with a potential $U(\phi)$ is governed,  in the \emph{Jordan frame}, by the following action\footnote{As usual, $M_P=1/\sqrt{8\pi G_N}\simeq 2.43\times 10^{18}$ GeV is the reduced Planck mass.}
\be
S=\int \d{^4 x}\,\sqrt{-g} \[\frac{M_P^2}{2} R +\frac{\xi}{2}  R \phi^2-\frac12(\del\phi)^2-U(\phi)\]~,
\ee
where indices are contracted with the metric $g_{\mu\nu}$, defined as
$\d s^2=-\d t^2 +a^2(t)\d x ^2$. Inflation can be conveniently studied in the {\it Einstein frame}, after performing a conformal transformation $g^E_{\mu\nu}=\Omega(\phi) g_{\mu\nu}$, with
$\Omega\equiv1+\xi\phi^2/M_P^2$ and canonically-normalizing the scalar field.  Up to a total derivative, the action takes the  familiar form 
\be
S=\int \d{^4 x}\, \sqrt{-g_E} \(\frac{M_P^2}{2}\, R_E -\frac12 g^{\mu\nu}_E \del_\mu\varphi\del_\nu\varphi- V\[\phi\(\varphi\)\]\)~,
\label{eq:ac}
\ee
where now $\varphi$ is the canonically-normalized inflaton,  related to the the original non-minimally coupled scalar field $\phi$ through 
\be
\(\frac{\d{\varphi}}{\d{\phi}}\)^2=\frac{1}{\Omega}+\frac32
M_P^2
\(\frac{\Omega'}{\Omega}\)^2~. 
\label{eq:jacq}
\ee
In terms of the original scalar field $\phi$,  the {\em physical potential} takes the simple form 
\be
V\[\varphi(\phi)\]={U(\phi)}/{\Omega^2(\phi)}\,. 
\label{eq:potential}
\ee
In the following, as previously stated, we shall focus on the simplest inflationary 
model. A generalization to other interesting inflationary
  scenarios, as for instance, the Higgs inflation model~\cite{Bezrukov:2007ep}, will
  be carried out elsewhere~\cite{inprep}. The simplest scenario is given by
the quadratic potential $U(\phi)=\frac12 m^2\phi^2$,  with a
non-vanishing coupling $\xi$. 
In order to derive the primordial scalar and tensor perturbation spectra within the non-minimally coupled $\phi^2$ theory, we shall make use of the slow-roll parameters\footnote{Here, we use the notation $\xi_{\textrm{SR}}(\varphi)$ to refer to the usual slow-roll parameter $\xi$, in order to avoid confusion with the non-minimal coupling to gravity $\xi$.}:
\bea
\epsilon\equiv \frac{M_P^2}{2} \(\frac{V_\varphi}{V}\)^2,
\eta\equiv M_P^2\frac{V_{\varphi\varphi}}{V}, 
\xi_{\textrm{SR}}\equiv M_P^4\frac{V_\varphi V_{\varphi\varphi\varphi}}{V^2}.
\label{eq:slowroll}
\eea
It is straightforward to derive the expressions for the spectral index of the primordial scalar perturbations $n_s\equiv 1+2\eta-6\epsilon $, its running $\alpha\equiv\d{n_s}/\d{\ln k} \equiv -24\epsilon^2+16\epsilon\eta-2\xi_{\textrm{SR}}$, 
and the tensor-to-scalar ratio $r \equiv 16\epsilon$ from the above slow-roll parameters\footnote{Notice that the expressions for both $n_s$ and $r$ are first-order in slow-roll, while $\alpha$ involves second order slow-roll terms. However, we have checked numerically that such second order corrections in slow-roll leave unchanged the constraints on the inflationary observables $(n_s, r)$. Therefore,  higher order slow-roll corrections can be safely neglected.}.

Within the slow-roll approximation, one can easily solve numerically the inflationary dynamics governed by the action \eq{eq:ac}. The number of e-folds is given by 
\be
N={1\over M_P}\int_{\varphi_{\rm end}}^{\varphi_\ast}
\frac{\d{\varphi}}{\sqrt{2 \epsilon(\varphi)}}={1\over
  M_P}\int_{\phi_{\rm end}}^{\phi_\ast} \frac{\d{\phi}}{\sqrt{2
    \epsilon(\phi)}} \left(\frac{\d{\varphi}}{\d{\phi}}\right)\,,
\label{eq:nefolds}
\ee
where $\epsilon(\phi)\equiv \frac{M_P^2}{2} \[V'(\phi)/V(\phi)\]^2$.

\begin{figure}[!t]
\vspace*{-0.5cm}
\includegraphics[width=0.5 \textwidth, height= 9.5cm]{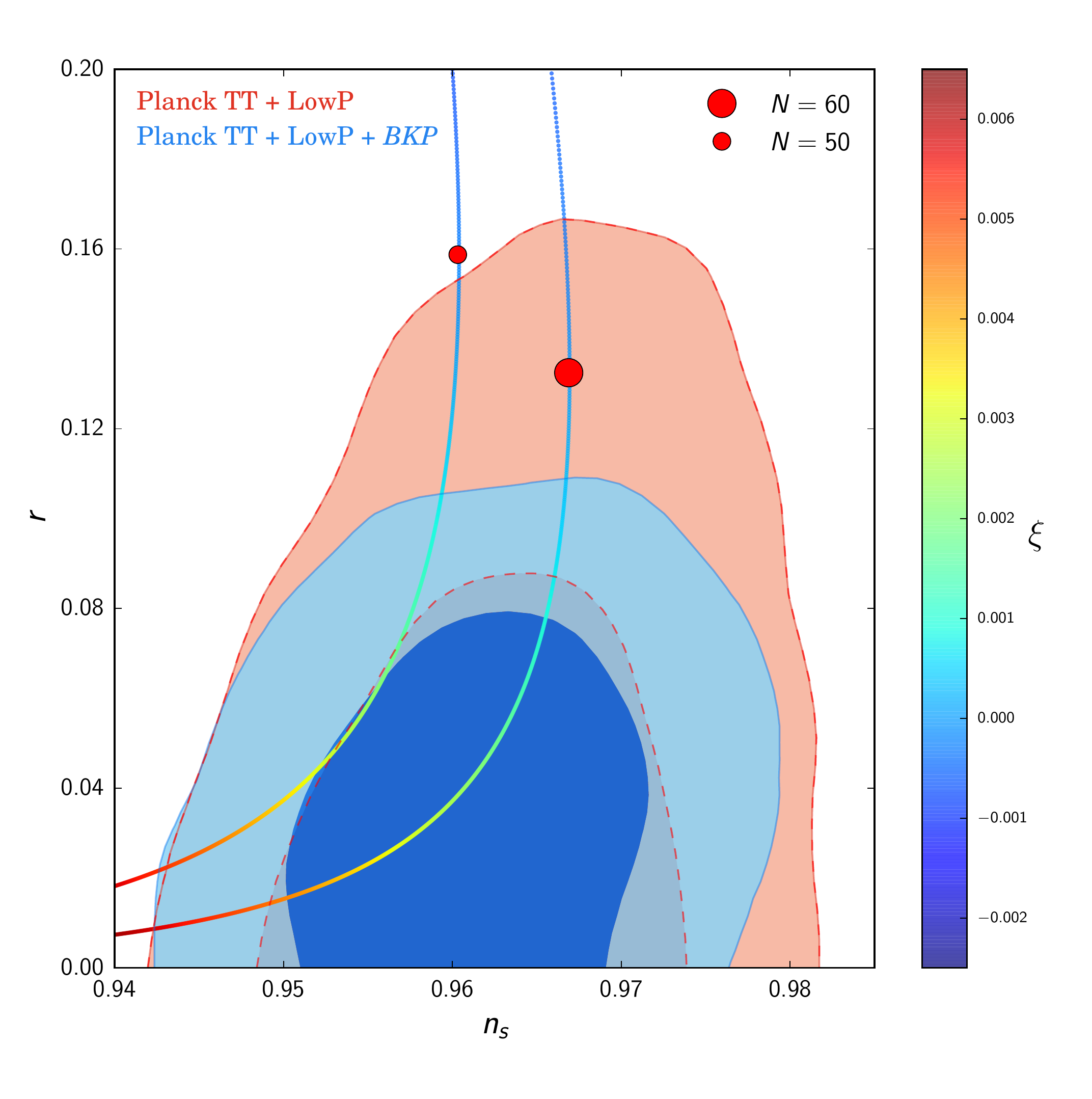}
\vspace*{-8mm}
\caption{Theoretical predictions for the chaotic model
  $V\propto\phi^2$ with a non-minimal coupling $\xi$ in the $(n_s,r)$
  plane for $N=50$ and $N=60$. The red circles represent the $\xi=0$ case,
  corresponding to the usual predictions of the chaotic inflationary
  scenario. We show as well the $68\%$ and  $95\%$ confidence level
  regions arising from the usual analyses in the $(n_s,r)$ plane using the various data combinations
  considered here.}
 \label{fig:xi}
\end{figure}

The inflationary theoretical predictions for the $N=50$ and $N=60$ cases are
depicted in Fig.~\ref{fig:xi}, in the $(n_s,r)$ plane, for both
positive and negative values of the coupling $\xi$. The case of
$\xi=0$ corresponds to the usual predictions of the chaotic
inflationary scenario, with $n_s=1-2/N \simeq 0.967$ ($n_s \simeq 0.96$) and $r =8/N\simeq
0.13$ ($r \simeq 0.16$) for $N=60$ ($N=50$), and it is represented  by
red circles. Notice that negative values of $\xi$ lead to a larger
tensor-to-scalar ratio. Positive values of $\xi$, on the other hand,
will reduce the tensor contribution, while also pushing $n_s$
significantly below scale invariance as $\xi$ increases. For instance,
for $\xi >0.002$ and $N=60$,  the scalar spectral index will always be smaller than the  observationally  preferred value $n_s\simeq 0.96$. 

The predicted running of the spectral index $\alpha$ is shown in  
Fig.~\ref{fig:alphar} as a function of the non-minimal coupling
$\xi$. In general, negative (positive) values of $\xi$ lead to
positive (negative) values of the running. Although the large positive
values of the running shown in Fig.~\ref{fig:alphar} are compatible
with the recent {\it Planck} 2015 constraints \cite{planckcosmo}, 
$\alpha=-0.0065\pm 0.0076$, they are nevertheless associated with
values of the tensor-to-scalar ratio $r>0.5$, which are  excluded
observationally.  The red circle in Fig.~\ref{fig:alphar} refers to
the $\xi=0$ case, corresponding to $\alpha=-2/N^2\simeq-0.00056$ for
$N=60$.
\squeezetable
\begin{table}[!b]
\begin{center}
\begin{tabular}{c|l|c}
\hline\hline
 Parameter & Physical Meaning  & Prior\\
\hline
$\omega_b\equiv\Omega_{b}h^2$ & Present baryon density.&$0.005 \to 0.1$\\
$\omega_c\equiv\Omega_{c}h^2$ &Present Cold dark matter density.&$0.01 \to 0.99$\\
$\Theta_s$ &$r_s/D_A(z_{\rm dec})$~\footnote{The parameter $\Theta_{s}$  is the ratio between the sound horizon $r_s$ and the angular diameter distance $D_A(z_{\rm dec})$ at decoupling $z_{\rm dec}$.}. & $0.5 \to 10$\\
$\tau$ &Reionization optical depth. &$0.01 \to 0.8$\\
$\ln{(10^{10} A_{s})}$& Primordial scalar amplitude.& $2.7 \to 4$\\
$\xi$  &Non-minimal coupling.&$ -0.002 \to  0.0065 $\\[1mm]
\hline\hline
\end{tabular}
\caption{Uniform priors for the cosmological parameters considered in the present analysis.}
\label{tab:priors}
\end{center}
\end{table}

\section{Observational constraints on $\xi$ in the quadratic
  inflationary model}

 In this paper, we restrict our numerical fits to  Cosmic Microwave Background (CMB) measurements. The inclusion of external data sets, such as Baryon Acoustic Oscillation
measurements, or a Hubble constant prior from the HST team
will not affect the constraints presented in the following. Our data sets are the {\it Planck} temperature data
(hereafter \emph{TT})~\cite{Ade:2013ktc,Planck:2013kta,Ade:2013tyw}, together with
the low-$\ell$ WMAP 9-year polarization likelihood, that includes
multipoles up to $\ell=23$, see Ref.~\cite{Bennett:2012zja} (hereafter \emph{WP}), and the
recent multi-component likelihood of the joint analysis of
BICEP2/\emph{Keck} Array and {\it Planck} polarization maps (hereafter \emph{BKP}), following
the data selection and foreground parameters of the fiducial
analysis presented in Ref.~\cite{Ade:2015tva}~\footnote{This fiducial analysis
assumes a tensor spectral index $n_T=0$, the $BB$ bandpowers of BICEP2/\!\!\!
\emph{Keck} Array and the $217$ and $353$~GHz bands of {\it Planck}, in the multipole range $20< \ell < 200$.}. However, variations of this fiducial
model will not change significantly the results presented here. 

These data sets are combined to constrain the cosmological model explored here, and described by the  parameters\footnote{Notice that the inflationary cosmology under study contains less parameters than the standard $\Lambda$CDM picture, as once the non-minimal coupling $\xi$ is fixed,  $n_s$, $r$ and $\alpha$ are fully determined, and are thus {\it derived} parameters.}:
\bea
  \{\omega_b,\omega_c, \Theta_s, \tau, \log[10^{10}A_{s}], \xi\}
\eea
\noindent
In Table \ref{tab:priors}, we summarize the definition as well as the priors on these parameters. We use the Boltzmann code {\tt CAMB}~\cite{camb} and 
the cosmological parameters are extracted from the data described 
above by means of a Monte Carlo Markov Chain (MCMC) analysis based on 
the most recent version of {\tt cosmomc}~\cite{Lewis:2002ah}. 
The constraints obtained on the non-minimal coupling $\xi$
are then translated into bounds on the usual inflationary parameters
$n_s$, $r$ and $\alpha$.
\begin{table*}
\begin{center}
\begin{tabular}{|c|cc|cc|}
\hline \hline
&\multicolumn{2}{c|}{Planck TT+WP}&\multicolumn{2}{c|}{BK+Planck TT+WP}\\[2mm] 
\hline 
$N$ & $60$ &$50$& $60$&$50$
\\[1mm]
$\xi$ & $0.0028_{-0.0025}^{+0.0023}$ & $0.0024_{-0.0023}^{+0.0023}$ &$0.0027_{-0.0022}^{+0.0023}$ & $0.0027_{-0.0019}^{+0.0020}$
\\[5mm]
\hline
\hline
$n_s$  & $0.958_{-0.011}^{+0.010}$ & $0.954_{-0.009}^{+0.007}$& $0.958_{-0.011}^{+0.009}$ & $0.953_{-0.009}^{+0.007}$
\\[5mm]
$r$ & $ 0.038_{-0.031}^{+0.051}$ &$0.063_{-0.048}^{+0.056}$ & $0.038_{-0.030}^{+0.039}$ & $0.053_{-0.037}^{+0.038}$\\[5mm]
$\alpha\equiv \d n_s/\d\ln k$& $-0.0005_{-0.0001}^{+0.0001}$ &$-0.0007_{-0.0001}^{+0.0001}$& $-0.0005_{-0.0001}^{+0.0001}$ &$-0.0007_{-0.0001}^{+0.0001}$\\[2mm]
\hline
\hline
\end{tabular}
\caption{\textbf{Inflationary constraints in the context of non-minimally coupled
    chaotic potential $\phi^2$}: The upper block of the table refers to
  the $95\%$~CL limits on the non-minimal coupling $\xi$ (the parameter varied in the MCMC analyses) from the two
  possible CMB data combinations used in this study, for both $N=60$
  and $N=50$. The lower block of the table contains the $95\%$~CL derived ranges of the inflationary parameters $n_s$, $r$
  and $\alpha$ from the limits of $\xi$ illustrated above, in the context of
  the non-minimally coupled chaotic potential $\phi^2$, for both $N=60$ and $N=50$.}
\label{tab:95cl}
\end{center}
\end{table*}

\begin{figure}[!t]
\hspace*{-0.5cm}
\includegraphics[width=0.47\textwidth]{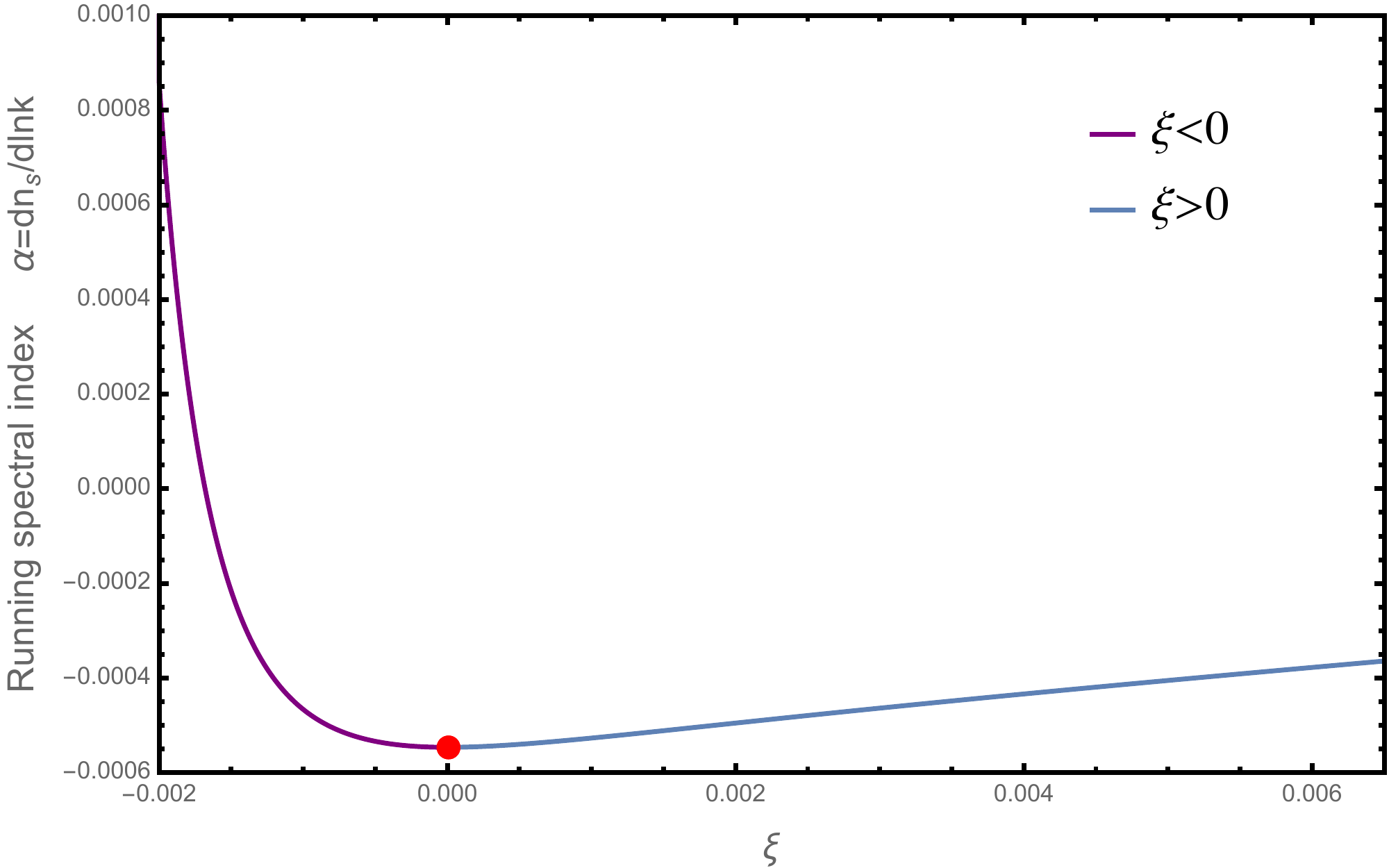}
 \caption{The running $\alpha$ as a function of the non-minimal coupling $\xi$.  The red circle represents the minimal coupling case $\xi=0$. }
\label{fig:alphar}
\end{figure}

\begin{figure*}[!t]
\hspace{-0.2cm}
\begin{tabular}{c c}
\includegraphics[width=0.48\textwidth]{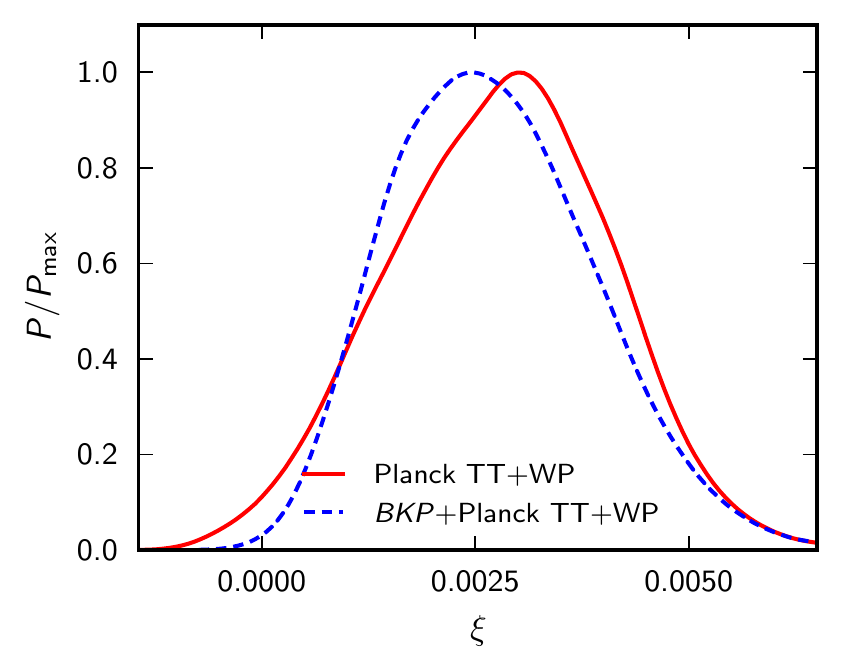}&~~~~ 
\includegraphics[width=0.48\textwidth]{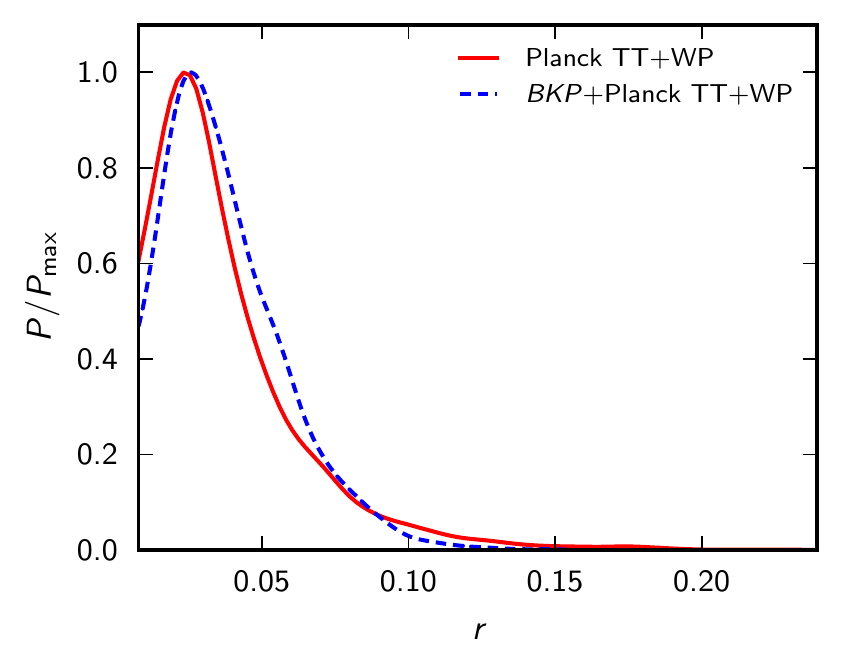}\\
\end{tabular}
 \caption{The left (right) panel shows the one-dimensional
   posterior probability distributions of the non-minimal coupling $\xi$ (the
tensor-to-scalar ratio $r$) in the context of a non-minimally coupled
    chaotic potential $\phi^2$, with $r$ a function of $\xi$, and therefore, a prediction within the model.}
\label{fig:nsr}
\end{figure*}

Table \ref{tab:95cl} shows the $95\%$~CL constraints on the parameter  $\xi$ as well as on the derived inflationary parameters $n_s$, $r$ and the running 
$\alpha$ arising from our numerical analyses using the two CMB data
combinations used here and assuming that $n_s$ and $r$ are univocally
determined by $\xi$ (for a fixed number of e-folds $N$, that we consider
to be either $60$ or $50$). For $N=60$, the preferred value of the non-minimal
coupling $\xi$ from {\it Planck} \emph{TT} plus \emph{WP}  
measurements is positive and slightly larger than the mean
value obtained when the cross-correlated polarized maps from BICEP2/\emph{Keck} and {\it Planck} (\emph{BKP}) experiments are included in the numerical analyses. This
preference for a slightly larger $\xi$ (and consequently, smaller $r$)
is clear from the one-dimensional posterior probability
distribution of $\xi$ shown in the left panel of Fig.~\ref{fig:nsr}. 
The mean value of $\xi=0.0028$ obtained from {\it Planck} \emph{TT} plus
\emph{WP} data is translated into a $95\%$~CL constraint of the tensor-to-scalar-ratio $r=0.038_{-0.031}^{+0.051}$, as can be seen from the right panel of Fig.~\ref{fig:nsr}.  When considering BICEP2/\emph{Keck} and {\it Planck} cross-spectra polarization data, the former constraint on the tensor-to-scalar ratio is very similar to the one quoted above. Concerning the running of
the spectral index, the two data combinations seem to have a
preference for a small negative running $\alpha=-0.0005$, associated to
 small values of $|\xi|$, as shown in Fig.~\ref{fig:alphar}.

Let us now comment on the sensitivity of our constraints to changes in
the number of e-folds $N$. Setting $N=50$ leads to different, though
almost insignificant, changes in the constraints obtained using the
two CMB data sets. The theoretically allowed regions in the $(n_s, r)$
plane as a function of $\xi$ for $N=50$ are indeed slightly different
from those corresponding to the $N=60$ case, see Fig.~\ref{fig:xi}. 
The net result is a smaller (larger) values of $n_s$ ($r$) than in the
$N=60$ case. The BICEP2/\emph{Keck} and {\it  Planck} cross-spectra 
polarization data yield a value $r=0.053_{-0.037}^{+0.038}$ for 
the tensor-to-scalar ratio in the context of the non-minimally coupled
$\phi^2$ model. On the other hand, the resulting central value for the 
scalar spectra index is only half a $\sigma$ away (towards smaller
values) from the corresponding one for $N=60$, as expected from the
theoretical predictions illustrated in Fig.~\ref{fig:xi}.

Figure~\ref{fig:xi} shows the $68\%$ and $95\%$~CL allowed regions in
the $(n_s,r)$ plane resulting from our MCMC analyses to {\it Planck} \emph{TT} plus \emph{WP} 
data  and to the combined \emph{BKP} in the usual $(n_s, r)$ plane, together with the theoretical
predictions for $N=50$ and $N=60$ for the non-minimally coupled
$\phi^2$ scenario.


\begin{figure}[!b]
\hspace*{-7mm}
\includegraphics[width=0.5\textwidth]{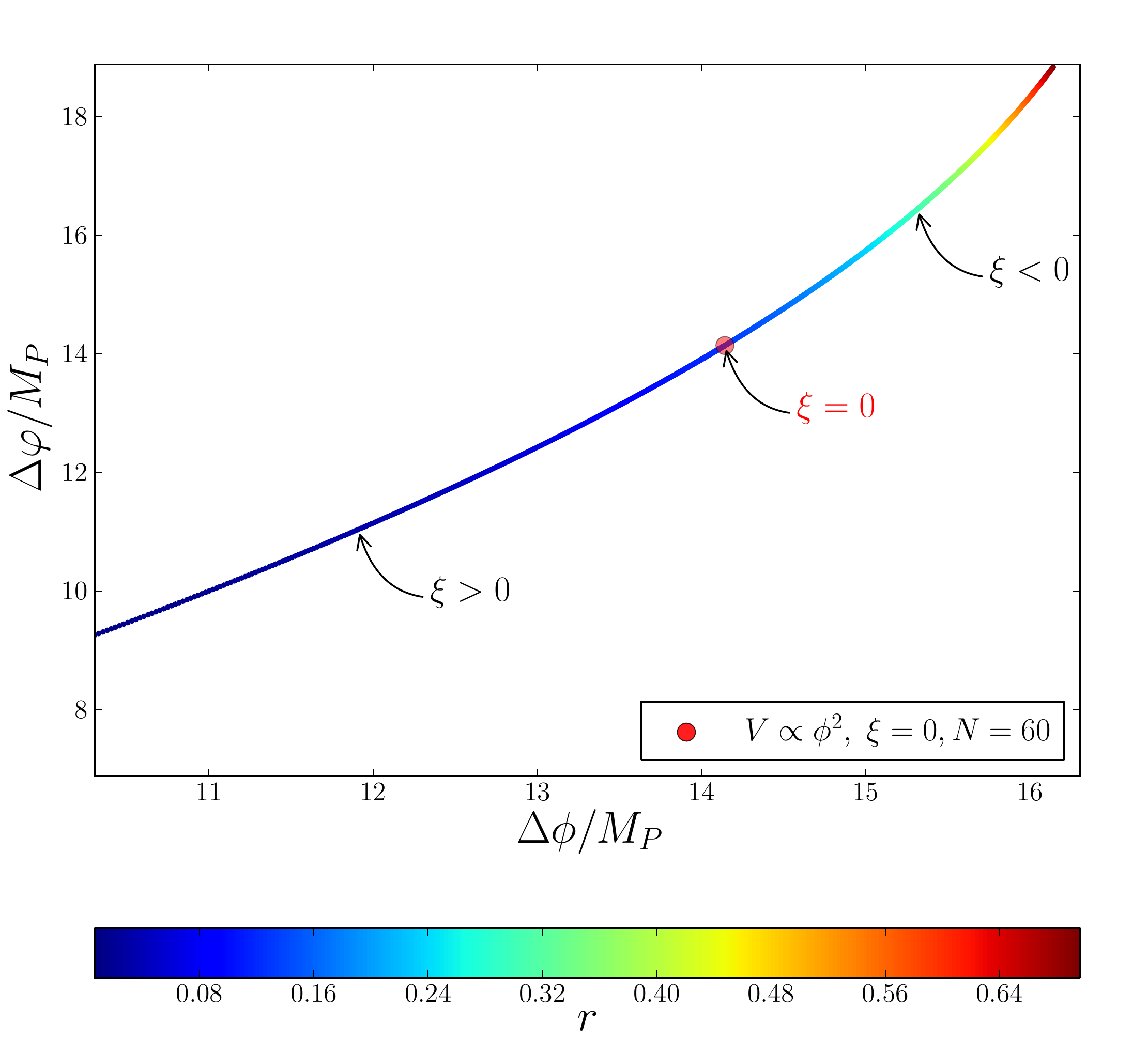}
\caption{Excursion of the canonically-normalized inflaton $\varphi$
  versus the one of the original scalar $\phi$. The magnitude of the
  tensor-to-scalar ratio is encoded in the curve through the color
  bar. Notice that, in both frames, large $r$ correlates with large excursions. }
\label{fig:deltaphi}
\end{figure}

To address the question of whether or not a non-minimal coupling $\xi$ is favoured by current CMB data, we compare the $\chi^2$ test
statistics function for the $\phi^2$ model in its minimally and
non-minimally coupled versions for $N=60$, albeit very similar results
are obtained for $N=50$. The $\chi^2$ for the case of
 {\it Planck} \emph{TT} plus \emph{WP} data, evaluated at
the best-fit-point of the $\phi^2$ model minimally coupled to gravity
is $\chi^2[\xi=0]=9812.8$. On the other hand, the non-minimally coupled version has a lower $\chi^2$ value at the best-fit-point due to the extra parameter $\xi$  introduced in the model, with  $\chi^2[\xi\ne 0]= 9806.8$. The difference between these two $\chi^2$ values is $\Delta\chi^2=6$, which, for a distribution
 of one degree of freedom, has a $p$-value of $0.014$, and is considered  as statistically significant. For the case of the combined \emph{BKP} likelihood, the difference between the test-statistics for
 the minimally coupled and non-minimally coupled $\phi^2$ models is
 $\Delta \chi^2=10$, which, for one degree of freedom, has a $p$-value of $0.0016$, and is considered as very statistically significant. Therefore, according to the most recent CMB data, the presence of a non-minimal coupling $\xi$ within the $\phi^2$ model is favoured at a significance equal or larger than $\sim 99\%$~CL.
 
Let us now turn to future constraints on $\xi$. Future
observations, as those expected from {\it PIXIE}~\cite{Kogut:2011xw}, {\it Euclid}~\cite{Laureijs:2011gra},  {\it COrE} \cite{Bouchet:2011ck}
and {\it PRISM}~\cite{Andre:2013afa},  
could be able to reach an accuracy of  
$\sigma_r=\sigma_{n_s-1}=10^{-3}$. With such precision, one could hope to test deviations from the quadratic potential~\cite{Creminelli:2014oaa}, as the one studied here,  by constructing quantities independent of $N$, up to subleading ${\mathcal O}(1/N^3)$ corrections. It is straighforward to get for our case,    
\be
n_s-1+\frac{r}{4}=-20\,\xi\,, 
\ee
at leading order both in slow-roll and $\xi$. If it turns out that nature had chosen  a very small value of $r$, future constraints on $\xi$ would be as strong as $\xi\lesssim 10^{-4}$; one order of magnitude stronger than the ones obtained in this analysis. Concerning the  running $\alpha$, it is interesting to note that futuristic observations like {\it SPHEREx} \cite{Dore:2014cca} with a forcasted error of $\sigma_\alpha= 10^{-3}$, will be able to falsify the present scenario.

Finally, it is also interesting to explore the impact of the
non-minimal coupling on the inflaton excursion.  It is well-know that
large values of the tensor-to-scalar ratio $r$, as those found by
previous BICEP2 measurements~\footnote{The joint BKP analysis
  finds however no evidence for primordial $B$-modes, but a robust
  limit of $r<0.12$ at  $95\%$ CL, see Ref.~\cite{Ade:2015tva}.}
\cite{bicep21,bicep22} yield large inflaton excursions $\phi \gg M_P$~\cite{Lyth:1996im, Boubekeur:2012xn, Efstathiou:2005tq, Verde:2005ff}, which are hard to understand in the context of a consistent effective field theory. In particular, successful inflation requires that higher order non-renormalizable operators, which are expected to be naturally present in the inflationary potential, are sufficiently suppressed.  A number of phenomenological studies have recently been devoted to address this problem \cite{Garcia-Bellido:2014eva, Barenboim:2014tca,Garcia-Bellido:2014wfa,Boubekeur:2014xva}.  
In Fig.~\ref{fig:deltaphi}, we plot the excursion of both $\phi$ and $\varphi$, together with the corresponding tensor-to-scalar ratio $r$. It is clear that the excursion of the canonically-normalized inflaton 
$\varphi$ is lowered for positive values of $\xi$ \ie
$\Delta\varphi<\Delta\phi$. However, this decrease is rather mild and
the excursion still takes on super-Planckian values for the
phenomenologically acceptable values of $\xi$. Conversely, negative
values  of $\xi$ lead  to an increase of the excursion of
$\varphi$. Figure~\ref{fig:deltaphi} also shows that super-Planckian
values of both $\phi$ and $\varphi$ are {\it still} associated with
large values of the tensor-to-scalar ratio $r$, in agreement with the
Lyth bound \cite{Lyth:1996im}. Thus, once a  small non-zero and
positive value of the coupling $\xi$ is turned on, both the inflaton
excursion and $r$ are slightly lowered, but without alleviating
completely the super-Planckian excursion problem. 

\section{Conclusions} 

A small, non-minimal coupling $\frac12 \xi
R\phi^2$ is expected to be present in the inflaton Lagrangian, and
modifies the inflationary predictions in an interesting way. Focusing
on the simplest quadratic potential scenario,  and using the very
recent joint analysis of BICEP2/\emph{Keck} Array and {\it Planck}
polarization maps, we found that a small, positive value of the
coupling $\xi$ is favoured at the $2\sigma$ level, assuming that nature has chosen the $\phi^2$ scenario
for the generation of primordial perturbations. If only {\it Planck}
\emph{TT} plus \emph{WP} data are used in the analyses, the
significance is milder. These conclusions have been obtained for a
number of e-foldings within the $N=50-60$ range. It would be interesting to see if upcoming $B$-modes
measurements can reinforce or weaken the statistical significance of these findings. In particular, it would be crucial to discriminate between the presence of a non-minimal coupling in the theory and other departures from the quadratic approximation.

\section{Acknowledgments}
OM is supported by PROMETEO II/2014/050, by the Spanish Grant FPA2011--29678 of the MINECO and by PITN-GA-2011-289442-INVISIBLES. LB and HR acknowledge financial support from PROMETEO II/2014/050.


\begin{thebibliography}{99}


\bibitem{Ade:2015tva}
  P.~A.~R.~Ade {\it et al.}  [BICEP2 and Planck Collaborations],
  ``A Joint Analysis of BICEP2/Keck Array and Planck Data,''
  [arXiv:1502.00612 [astro-ph.CO]].

\bibitem{planckcosmo} [Planck Collaboration],
  ``Planck 2015 results. XIII. Cosmological parameters,''
  arXiv:1502.01589 [astro-ph.CO].
\bibitem{planckng}
P.~A.~R.~Ade {\it et al.}  [Planck Collaboration], Planck 2015 results. XVII. ``Constraints on primordial non-Gaussianity,'' 
arXiv:1502.01592 [astro-ph.CO].

\bibitem{planckinf}
P.~A.~R.~Ade {\it et al.}  [Planck Collaboration],
``Planck 2015. XX. Constraints on inflation,''
arXiv:1502.02114 [astro-ph.CO].
  
\bibitem{Buchbinder:1992rb}
  I.~L.~Buchbinder, S.~D.~Odintsov and I.~L.~Shapiro,
  {\em Effective action in quantum gravity,}
  Bristol, UK: IOP (1992) 413 pp. 

\bibitem{Salopek:1988qh} 
  D.~S.~Salopek, J.~R.~Bond and J.~M.~Bardeen,
  Phys.\ Rev.\ D {\bf 40}, 1753 (1989).

\bibitem{Futamase:1987ua}
  T.~Futamase and K.~i.~Maeda,
  Phys.\ Rev.\ D {\bf 39} (1989) 399.

 
 \bibitem{Fakir:1990eg} 
  R.~Fakir and W.~G.~Unruh,
  Phys.\ Rev.\ D {\bf 41}, 1783 (1990).
 
 \bibitem{Kaiser:1994vs} 
  D.~I.~Kaiser,
  Phys.\ Rev.\ D {\bf 52}, 4295 (1995)
  [astro-ph/9408044].
 
 \bibitem{Komatsu:1999mt} 
  E.~Komatsu and T.~Futamase,
  Phys.\ Rev.\ D {\bf 59}, 064029 (1999)
  [astro-ph/9901127].
 
 \bibitem{Hertzberg:2010dc} 
  M.~P.~Hertzberg,
  JHEP {\bf 1011}, 023 (2010)
  [arXiv:1002.2995 [hep-ph]].

\bibitem{Okada:2010jf}
  N.~Okada, M.~U.~Rehman and Q.~Shafi,
  Phys.\ Rev.\ D {\bf 82} (2010) 043502
  [arXiv:1005.5161 [hep-ph]].

  \bibitem{Linde:2011nh} 
  A.~Linde, M.~Noorbala and A.~Westphal,
  JCAP {\bf 1103}, 013 (2011)
  [arXiv:1101.2652 [hep-th]].
  
\bibitem{Kaiser:2013sna}
  D.~I.~Kaiser and E.~I.~Sfakianakis,
  Phys.\ Rev.\ Lett.\  {\bf 112} (2014) 1,  011302
  [arXiv:1304.0363 [astro-ph.CO]].
\bibitem{Chiba:2014sva} 

  T.~Chiba and K.~Kohri,
  PTEP {\bf 2015}, no. 2, 023E01
  [arXiv:1411.7104 [astro-ph.CO]].

\bibitem{Pallis:2014cda} 
  C.~Pallis and Q.~Shafi,
  JCAP {\bf 1503}, no. 03, 023 (2015)
  [arXiv:1412.3757 [hep-ph]].

\bibitem{Tsujikawa:2013ila}
  S.~Tsujikawa, J.~Ohashi, S.~Kuroyanagi and A.~De Felice,
  Phys.\ Rev.\ D {\bf 88} (2013) 2,  023529
  [arXiv:1305.3044 [astro-ph.CO]].
  
 \bibitem{Linde:1983gd}
  A.~D.~Linde,
  Phys.\ Lett.\ B {\bf 129} (1983) 177.
  
\bibitem{Bezrukov:2007ep} 
  F.~L.~Bezrukov and M.~Shaposhnikov,
  Phys.\ Lett.\ B {\bf 659}, 703 (2008)
  [arXiv:0710.3755 [hep-th]].
  
\bibitem{inprep} 
  L.~Boubekeur, E.~Giusarma, O.~Mena and H.~Ram\'irez, \\
  {\it In preparation}. 


\bibitem{Ade:2013ktc} 
  P.~A.~R.~Ade {\it et al.}  [Planck Collaboration],
  arXiv:1303.5062 [astro-ph.CO].
  
\bibitem{Planck:2013kta} 
  P.~A.~R.~Ade {\it et al.}  [Planck Collaboration],
  arXiv:1303.5075 [astro-ph.CO].
  
\bibitem{Ade:2013tyw} 
  P.~A.~R.~Ade {\it et al.}  [Planck Collaboration],
  arXiv:1303.5077 [astro-ph.CO].
 \bibitem{Bennett:2012zja} 
  C.~L.~Bennett {\it et al.}  [WMAP Collaboration],
  Astrophys.\ J.\ Suppl.\  {\bf 208}, 20 (2013)
  [arXiv:1212.5225 [astro-ph.CO]].

\bibitem{camb}
  A.~Lewis, A.~Challinor and A.~Lasenby,
  Astrophys.\ J.\  {\bf 538}, 473 (2000)
  [arXiv:astro-ph/9911177].
%
\bibitem{Lewis:2002ah}
  A.~Lewis and S.~Bridle,
  Phys.\ Rev.\  D {\bf 66}, 103511 (2002)
  [arXiv:astro-ph/0205436].


\bibitem{Kogut:2011xw}
  A.~Kogut, D.~J.~Fixsen, D.~T.~Chuss, J.~Dotson, E.~Dwek, M.~Halpern, G.~F.~Hinshaw and S.~M.~Meyer {\it et al.},
  JCAP {\bf 1107} (2011) 025
  [arXiv:1105.2044 [astro-ph.CO]].




\bibitem{Laureijs:2011gra}
  R.~Laureijs {\it et al.}  [EUCLID Collaboration],
  ``Euclid Definition Study Report,''
  arXiv:1110.3193 [astro-ph.CO].
  



  \bibitem{Bouchet:2011ck}
  F.~R.~Bouchet {\it et al.}  [COrE Collaboration],
  ``COrE (Cosmic Origins Explorer) A White Paper,''
  arXiv:1102.2181 [astro-ph.CO].

\bibitem{Andre:2013afa}
  P.~Andre {\it et al.}  [PRISM Collaboration],
  ``PRISM (Polarized Radiation Imaging and Spectroscopy Mission): A White Paper on the Ultimate Polarimetric Spectro-Imaging of the Microwave and Far-Infrared Sky,''
  arXiv:1306.2259 [astro-ph.CO].



\bibitem{Creminelli:2014oaa}
  P.~Creminelli, D.~L\'opez Nacir, M.~Simonovi\'c, G.~Trevisan and M.~Zaldarriaga,
  Phys.\ Rev.\ Lett.\  {\bf 112} (2014) 24,  241303
  [arXiv:1404.1065 [astro-ph.CO]].





\bibitem{Dore:2014cca}
  O.~Dor\'e, J.~Bock, P.~Capak, R.~de Putter, T.~Eifler, C.~Hirata, P.~Korngut and E.~Krause {\it et al.},
  ``SPHEREx: An All-Sky Spectral Survey,''
  arXiv:1412.4872 [astro-ph.CO].


\bibitem{bicep21}
P.~A.~R.~Ade {\it et al.}  [BICEP2 Collaboration],
  Phys.\ Rev.\ Lett.\  {\bf 112} (2014) 241101. 
  [arXiv:1403.3985 [astro-ph.CO]].
 \bibitem{bicep22}
P.~A.~R.~Ade {\it et al.}  [BICEP2 Collaboration],
Astrophys.\ J.\  {\bf 792} (2014) 62
[arXiv:1403.4302 [astro-ph.CO]].



  
\bibitem{Lyth:1996im}
  D.~H.~Lyth,
  Phys.\ Rev.\ Lett.\  {\bf 78} (1997) 1861
  [hep-ph/9606387].
  



\bibitem{Boubekeur:2012xn} 
  L.~Boubekeur,
  Phys.\ Rev.\ D {\bf 87} (2013) 6,  061301 
   [arXiv:1208.0210 [astro-ph.CO]].
  
\bibitem{Efstathiou:2005tq}
  G.~Efstathiou and K.~J.~Mack,
  JCAP {\bf 0505} (2005) 008
  [astro-ph/0503360].
  

\bibitem{Verde:2005ff}
  L.~Verde, H.~Peiris and R.~Jimenez,
  JCAP {\bf 0601} (2006) 019
  [astro-ph/0506036].
  
  
\bibitem{Garcia-Bellido:2014eva} 
  J.~Garcia-Bellido, D.~Roest, M.~Scalisi and I.~Zavala,
  JCAP {\bf 1409}, 006 (2014)
  [arXiv:1405.7399 [hep-th]].
  
 
 \bibitem{Barenboim:2014tca} 
  G.~Barenboim and O.~Vives,
  arXiv:1405.6498 [hep-ph].

\bibitem{Garcia-Bellido:2014wfa} 
  J.~Garcia-Bellido, D.~Roest, M.~Scalisi and I.~Zavala,
  Phys.\ Rev.\ D {\bf 90}, no. 12, 123539 (2014)
  [arXiv:1408.6839 [hep-th]].
  
\bibitem{Boubekeur:2014xva} 
  L.~Boubekeur, E.~Giusarma, O.~Mena and H.~Ram\'irez,
  arXiv:1411.7237 [astro-ph.CO].
  
\end{thebibliography}
\end{document}